\documentclass[preprint]{aastex}
\slugcomment{Accepted for publication in AJ}
\shorttitle{Restorative extraction of spectra}
\shortauthors{Walsh \& Lucy}
\begin{document}
\title{Iterative techniques for the decomposition of long-slit
spectra}

\author{L. B. Lucy}
\affil{Astrophysics Group, Imperial College London, Blackett Laboratory, 
Prince Consort Road, London SW7 2BZ, UK}
\email{l.lucy@ic.ac.uk}

\and

\author{J. R. Walsh}
\affil{Space Telescope European Coordinating Facility, European Southern
Observatory, Karl-Schwarzschild Strasse 2, D-85748 Garching, Germany}
\email{jwalsh@eso.org}
\begin{abstract}
Two iterative techniques are described for decomposing a long-slit 
spectrum into the individual spectra of the point sources along the 
slit and the spectrum of the underlying background. One technique 
imposes the strong constraint that the spectrum of the background is
spatially-invariant; the other relaxes this constraint. Both techniques 
are applicable even when there are numerous overlapping point sources 
superposed on a structurally-complex background. The techniques
are tested on simulated as well as real long-slit data from the ground and 
from space. 
\end{abstract}

\section{Introduction}
Long-slit spectroscopy can be used to advantage in the
investigation both of point sources and of extended emission. But for
optimum results, reduction packages of some sophistication are required since
the observed spectrum comprises the superposed spectra of all objects
falling within the slit.
Thus, for a point source, the spectrum of the underlying extended emission
and the spectra of any nearby
point sources
need to be subtracted in order to extract the uncontaminated 1-D spectrum of
the target. Correspondingly, for an extended object, the spectra of any 
superposed point sources must be subtracted in order to extract the
uncontaminated 2-D spectrum of the target.   

The subtraction tasks described above are clearly straightforward if
the intensity of the extended emission has an approximately linear variation
along the slit and if point sources are well separated. For such cases,
standard IRAF procedures are applicable and not easily improved upon. But
these same procedures must be expected to perform poorly when the observed
field is not so simple. Accordingly, the aim of this investigation is to
develop procedures for treating such cases. Examples which require
more complex treatment include spectra of star clusters, planetary
nebulae in Local Group Galaxies, H~II regions in spiral arms,
jets in young stars and galaxies, galactic nuclei and gravitational lenses.

Software that performs the above subtraction tasks in the
more challenging circumstances where the extended emission is structurally
complex and where the broadened spectra of point sources overlap are described
and tested in this paper. Of course, given the errors associated with real
data, no procedure can carry out these extraction tasks perfectly. Thus,
some cross contamination will always remain in the extracted spectra of
overlapping point sources as also between the extracted spectrum of a point
source and that of the underlying extended emission. Nevertheless, procedures
that perform such extractions with close to the optimum achievable reliability
are clearly desirable.

Section 2 introduces two channel restoration in imaging and points
out its relevance for the decomposition of long-slit spectra.
Section 3 presents the iterative technique for a homogeneous 
background and section 4 for the general case of an inhomogeneous 
background. Section 5 presents 
simulations and applications of both methods to real data. As 
this investigation was underway, other authors have presented 
their treatments of essentially the same problem and a comparison 
between these various 
approaches is given in Section 6. The software implementation of 
the two new methods is briefly described in Section 7 and the 
conclusions in section 8. 

\section{Two channel restorations}

The two iterative schemes for decomposing long-slit
spectra, to be described in Sections 3 and 4,
are variants of a previously-developed two-channel restoration
procedure for astronomical images. Accordingly, we first recall salient
aspects of this earlier work and identify an important innovation 
adopted here.  

The problem of extracting reliable spectra of point sources and
background
from a long-slit spectrum is technically similar to that of
achieving high photometric accuracy when restoring an {\em image}
containing point sources superposed on extended emission.
For this latter problem, Richardson-Lucy (RL) ((\citet{rich72}, 
\citet{lucy74}) or MaxEnt restorations are not
satisfactory because of the
ringing that occurs in the background close to the point sources.

Following earlier work by \citet{frie78}, \citet{holu94b} and
\citet{lucy94} adopted a two channel approach to overcome this problem. 
All objects that the investigator deems to be point
sources are allocated to one channel and all remaining emission to the 
second channel. The merit of this approach is that, rather than
being presented with the impossible task of discovering delta functions from
noisy and band-limited data, the restoration procedure
is {\em told} which objects are to be restored as delta functions.
Nevertheless, in order for the iterative procedure
to find the desired
solution, it also proves necessary to prevent the second channel
from partially modelling the point sources as sharp peaks in the extended 
emission. This
is accomplished by limiting the resolution of the second channel.

In the Lucy-Hook scheme \citep{holu94b}, the angular resolution of $\psi$,
the model of the
extended emission, was limited by adding an entropic regularization term
to the objective function to be maximized. This term penalizes differences  
between $\psi$ and $<\psi>$, the convolution of $\psi$ with an appropriate
resolution kernel. Although effective, this procedure has the
disadvantage of introducing a free parameter $\alpha$, the 
constant controlling the
relative importance of the regularization term in the objective function.
An implementation of the Lucy-Hook scheme for long-slit spectra
has been described by \citet{wals97} and applied \citep{wals01}.

In this paper, the required
resolution limit is instead imposed by defining the restored background
$\psi$ to be the convolution
of a non-negative auxiliary function $\chi$ with a resolution
kernel $R$. Clearly, $R$ itself then represents the narrowest
feature that
can appear in $\psi$. Accordingly, if the FWHM of $R$ exceeds that of
the PSF, the convolution of $\psi$ with the PSF cannot fit
the observed profile of a point source, which must therefore be modelled in
the first channel as a delta function of appropriate amplitude.

As with images, an astronomer reducing a long-slit
spectrum will often be confident in identifying some objects as point
sources and will wish to benefit from the improved reduction that can result
from providing such input. 

\section{Homogeneous background}

The two techniques described in this paper differ only in their
assumptions about the background. In this section, we make the strong,
restrictive assumption that the background source is homogeneous. Thus,
although the intensity of the background emission may vary along
the slit, its spectrum does not. An example is a long-slit spectrum in the
core of a globular cluster: numerous faint members then form a smooth
background whose spectrum to a good approximation is 
spatially-invariant because of orbit-mixing.

\subsection{Simplification}

To simplify the description of the scheme, we assume that the 
observed spectrum $\tilde{\phi}(\lambda,y)$ is a
continuous function of the two 
variables $\lambda$ and $y$, where $\lambda$ denotes wavelength and $y$ is the
spatial coordinate perpendicular to the direction of dispersion. We further
suppose that the spectra of the point sources have $\lambda$ -independent
centroids $y_{n}$ - i.e., they are dispersed horizontally.
Of course the actual code works with pixellated data and can correct
for the curvature of the point-source spectra.

\subsection{Model}

Let $f_n(\lambda)$ be the spectrum of the $n$th point source
and let $F(\lambda)$ be the spatially-invariant spectrum of the homogeneous
background whose
normalized spatial profile is $\psi(\eta)$. Then, if $P(y-\eta,\lambda)$
is the spatial distribution of a point source at $y=\eta$ due to
instrumental broadening and seeing (for a ground-based telescope), the two-
dimensional model
for the observed spectrum is
\begin{equation}
  \phi(\lambda,y)= F(\lambda) \theta(y,\lambda)+ \sum_{n} f_{n}(\lambda) P(y-y_{n},\lambda)
\end{equation}
where
\begin{equation}
 \theta(y,\lambda)=\int \psi(\eta) P(y-\eta,\lambda)d\eta
\end{equation}
is the broadenened and therefore $\lambda$-dependent profile of the
background.

Now, as discussed in Sec. 2, the model as defined by Eqs. (1)
and (2) is not yet satisfactory since a point source (especially one 
with a spectrum similar to that of the background) can be partially modelled
as a peak in $\psi(\eta)$. To eliminate this near indeterminacy, we impose a
resolution limit on $\psi$ by writing
\begin{equation}
 \psi(\eta)=\int \chi(\zeta) R(\eta-\zeta) d\zeta
\end{equation}
where $\chi(\zeta)$ is a non-negative normalized auxiliary function and 
$R(\eta-\zeta)$ is a non-negative normalized kernel function. If $R$
is a bell-shaped function with FWHM significantly greater than that of $P$,
then point-like peaks cannot appear in $\psi(\eta)$. By thus preventing the
point-sources from partially contaminating our model of the background,
we simultaneously improve the accuracy of the extracted point-source
spectra.        

Eqs. (1)-(3) define the model used to decompose a long-slit
spectrum under the assumption that the background is homogeneous. In fitting
this model to an observed spectrum $\tilde{\phi}(\lambda,y)$, the unknowns
to be estimated are the spectra $f_{n}(\lambda)$ and $F(\lambda)$ and the
auxiliary function $\chi(\zeta)$. Note that the spatial profile of the 
background $\psi(\eta)$ is not a basic unknown but is obtained incidentally
from $\chi(\zeta)$ via Eq.(3). 

The point-spread function $P(y-\eta,\lambda)$ and the resolution
kernel $R$ are user-supplied and thus treated as known functions in 
the iterative fitting procedure described below. The centroids $y_n$ 
of the point-source spectra are also user-supplied and are held 
fixed during the iterations. These $y_n$ can be input by the user or
determined by cross-correlation with the point-spread function.

\subsection{Iterative technique}

We now estimate the unknown functions $f_{n}$, $F$ and $\chi$  by
iteratively improving the fit of
$\phi$ to the observed spectrum $\tilde{\phi}$.
This iterative technique, which is based on the RL algorithm
(Richardson 1972; Lucy 1974),
alternates between improving the spectra $f_{n}(\lambda)$ and
$F(\lambda)$ and improving the auxiliary function
$\chi(\zeta)$. 

The scheme is initiated with flat, featureless estimates for
the unknown functions. Thus we take $f_{n}(\lambda) = F(\lambda) =1$
and set the normalized auxiliary
function $\chi(\zeta)= 1/Y$, where $Y$ is the spatial extent of the spectrum.

\subsubsection{Correcting the spectra}

After initiation (iteration $i=1$) or after a failed test of
convergence ($i>1$),
improved
estimates of the spectra $f_{n}(\lambda)$ and $F(\lambda)$ are obtained as
follows:\\

1) With the current estimate of the auxiliary function $\chi(\zeta)$,
the profile of the
background $\psi(\eta)$ is computed from Eq.(3).\\

Following this initial calculation, the remaining steps are carried
out at each discrete $\lambda$ in the observed spectrum.\\ 

2) The broadened profile of the
background $\theta(y,\lambda)$ is computed from Eq. (2).

3) With the current estimates of $f_{n}$ and $F$,
the $y$-dependence of $\phi(\lambda,y)$ is derived from
Eq. (1) 

4) Improved values of the amplitudes $f_{n}$ are obtained by
applying the correction factors
\begin{equation}
 c_{n}=\int \frac{\tilde{\phi}(\lambda,y)}{\phi(\lambda,y)}P(y-y_{n},\lambda)dy
\end{equation}

5) An improved value of $F$ is obtained by applying the correction
factor
\begin{equation}
 C =\int \frac{\tilde{\phi}(\lambda,y)}{\phi(\lambda,y)}\theta(y,\lambda)dy
\end{equation}

\vspace{0.4cm}
When the above correction factors have been applied at all
$\lambda$'s, the improved spectra
are input to this iteration's next stage - Sec.3.3.2 -wherein $\chi(\zeta)$ is
improved. 

\subsubsection{Correcting the auxiliary function}

Since the background is here assumed to have a spatially-invariant
spectrum, the profile function
$\psi(\eta)$ and the auxiliary function $\chi(\zeta)$ from which it is
derived are $\lambda$-independent. In principle, therefore, these functions
could be estimated from the data at a single $\lambda$, but at the price 
of then neglecting information at other $\lambda$'s. To avoid this, we 
project the
model and the data onto the $y$-axis - i.e., we integrate over $\lambda$.

Let $\Phi(y)$ denote the integral of  
$\phi(y,\lambda)$. Then, from Eq.(1),
\begin{equation}
 \Phi(y)= \Phi_{b}(y) + \sum_{n} \Phi_{n}(y)
\end{equation}
where $\Phi_{b}(y)$ and $\Phi_{n}(y)$ are the
corresponding integrals of the broadened background
$F(\lambda) \theta(y,\lambda)$ and of the $n$th broadened point source
$f_{n}(\lambda) P(y-y_{n},\lambda)$, respectively.

Now, using Eqs (1),(2) and (5), we can relate
$\Phi_{b}(y)$ to the auxiliary function 
$\chi(\zeta)$. The result is the integral equation
\begin{equation}
 \Phi_{b}(y)= {\cal F} \int \chi(\zeta) Q( y - \zeta) d \zeta
\end{equation}
where $\cal F$ denotes the integral of $F(\lambda)$ and
\begin{equation}
 Q(y-\zeta) = \frac{1}{\cal F} \int\!\! \int F(\lambda) P(y-\eta,\lambda) R(\eta-  \zeta) d \lambda d \eta
\end{equation}
Note that $Q(y-\zeta)$ is a normalized non-negative function.

Having derived this basic integral equation, we can now describe
the steps that result in an improved estimate for the auxiliary function
$\chi(\zeta)$.\\

1) The effective broadening function $Q(y-\zeta)$ 
defined by Eq. (8) is computed.

2)  The integrated model $\Phi(y)$ is computed from Eqs. (6) 
and (7).

3) An improved $\chi(\zeta)$ is obtained by applying the
correction factor 
\begin{equation}
 {\cal C}(\zeta) = \int \frac{\tilde{\Phi}(y)}{\Phi(y)} Q(y-\zeta) dy
\end{equation}
where $\tilde{\Phi}(y)$ denotes the $\lambda$-integrated observed spectrum
$\tilde{\phi}(\lambda,y)$. 

\vspace{0.4cm}
When the above correction has been made, control moves
to this iteration's next stage wherein convergence is
tested. 

\subsubsection{Test of convergence}

As measures of the corrections made in Sec. 3.3.1 to the spectra,
we compute
\begin{equation}
 \delta_{n} = \int |\Delta f_{n}| d \lambda / \int f_{n} d \lambda
\end{equation}
for each point source, and the corresponding quantity $\delta_{F}$
for the background. The symbol $\Delta$ is the difference in value between
successive iterations. 

Similarly, we assess the corrections made in Sec. 3.3.2 to     
the normalized function $\chi(\zeta)$ by computing the quantity
\begin{equation}
 \delta_{\chi} = \int |\Delta \chi| d \zeta
\end{equation}

If {\em any} of the quantities $\delta_{n}$, $\delta_{F}$ or
$\delta_{\chi}$ exceeds an appropriately small quantity
$\epsilon$, control returns to Sec. 3.3.1 in order to carry out the
$(i+1)$-th iteration. 

\subsubsection{Output}

If the above convergence test is satisfied, the reduction is finished
and it remains only to output the quantities of scientific interest. These
are the spectra  $f_{n}(\lambda)$ and 
$F(\lambda)$ and the normalized profile of the background $\psi(\eta)$.  
For some problems, the spectra of some or all of the point sources will be of
primary interest; in others, the spectrum of the background; in yet others,
the spatial profile of the background. 

In addition, the two-dimensional residual spectrum 
$\phi(\lambda,y)-\tilde{\phi}(\lambda,y)$ is computed and displayed. Ideally, 
these residuals will be entirely attributable to statistical fluctuations in
the observed spectrum $\tilde{\phi}(\lambda,y)$. But if some residuals are
significant, a flawed or incomplete reduction is indicated - see comment f)
below. Of particular interest in this regard is the use of this residual
spectrum to discover evidence that the background's spectrum is not
spatially-invariant and to identify the astrophysical source - e.g., nebular
emission. Note that a weak departure from spatial invariance
is likely to be {\em }less evident if the spectrum were reduced
without this assumption (Sec. 4) and then such a departure sought by
inspecting the two-dimensional background spectrum $F(\lambda,\eta)$ that
is part of the output in the inhomogeneous case.  

The auxiliary function $\chi(\zeta)$ is not part of the standard 
output. In particular, this function does not provide a better model of the
background's profile than does $\psi(\eta)$.    

\subsection{Comments}

The following comments are intended to clarify various aspects of this
iterative scheme:\\

a) The correction factors defined by Eqs. (4) and (5) are
obtained from the RL algorithm by noting that
the spectra $f_{n}(\lambda)$ and $F(\lambda)$ are spatially
broadened into their respective contributions to the observed spectrum
according to the functions $P(y-y_{n},\lambda)$ and $\theta(y,\lambda)$.
Similarly, the correction factor defined by Eq. (9)
follows from noting that $\chi(\zeta)$ is broadened into $\Phi_{b}(y)$
according to the function $Q(y-\zeta)$. Note that all these correction factors
are unity if the model exactly matches the observed spectrum.

b) The scheme does not correct the extracted spectra
$f_{n}(\lambda)$ and $F(\lambda)$ for instrumental broadening in the
$\lambda$- direction, and
thus they are not affected by deconvolution artifacts. Accordingly, the
subsequent
analysis of one of these point-source spectra can be carried out exactly as if
it were the spectrum of an isolated point source  
extracted with a conventional package. Similarly, the spectrum of the
background can be analysed as if the observation was of a source not 
contaminated by point sources. 

c) In conventional applications of the RL algorithm, it
is advisable to limit the number of iterations in order to avoid fitting
statistical fluctuations in the data (Lucy 1974). Here this is prevented
by the resolution limit imposed by the kernel $R$, and so
the iterations can be continued to convergence.

d) In the software package, the unknown functions $f_{n}$, $F$
and $\chi$ are represented as finite vectors. The converged solution
obtained with the above iteration scheme effectively determines the elements
of these vectors from the Principle of Maximum Likelihood (Lucy 1974).

e) Statistical fluctuations in the extracted spectra $f_{n}(\lambda)$ and $F(\lambda)$
are not reduced by imposing a resolution limit in the $\lambda$-direction
nor by restricting the number of iterations.
Instead, noise in the $f_{n}$ is reduced by an effective weighted co-addition
in the spatial direction using the
broadening function $P(y-y_{n},\lambda)$ (cf Horne 1986; Robertson 1986).
Similarly, noise
in $F$ is reduced by an effective weighted co-addition using the   
broadening function $\theta(y,\lambda)$.  
  
f) The model $\phi(\lambda,y)$ obtained with the above
iterative scheme will not necessarily fit the observed spectrum 
$\tilde{\phi}(\lambda,y)$ to within errors. Significant residuals will remain
if the assumptions underlying the model are violated. In particular, the
background's spectrum may not be spatially-invariant. Also the
user-supplied broadening function $P(y-y_{n},\lambda)$ may be inaccurate. 
 
Point sources not allocated to the first channel will also give
significant residuals since, by design, the resolution kernel $R$ then
prevents a close fit. In this case, a second reduction with additional 
designated point sources is clearly called for.
A further cause of significant residuals would be a
marginally-resolved source with FWHM less than that of $R$.             

\section{Inhomogeneous background}

In this section, we relax the constraint that the background is
homogeneous. Thus the general case is now treated where the background's
spectrum varies with position along the slit.

\subsection{Model}

If the background is {\em in}homogeneous, its
spectrum  $F(\eta,\lambda)$ is {\em not} separable into the product  
$\psi(\eta)F(\lambda)$ as assumed in Sec. 3.2. Accordingly, 
in this case, the model for the two-dimensional spectrum is

\begin{equation}
  \phi(\lambda,y)= \omega(\lambda,y)+ \sum_{n} f_{n}(\lambda) P(y-y_{n},\lambda)
\end{equation}
where
\begin{equation}
 \omega(\lambda,y)=\int F(\lambda,\eta) P(y-\eta,\lambda)d\eta
\end{equation}
is the broadened spectrum of the background.

Now, as in Sec. 3.2, we must prevent point sources from  
being partially modelled
as peaks in $F(\lambda,\eta)$. To eliminate this possibility, we again
impose a limit on the spatial resolution of the background by writing
\begin{equation}
 F(\lambda,\eta)=\int \chi(\zeta,\lambda) R(\eta-\zeta) d\zeta
\end{equation}
where $\chi(\zeta,\lambda)$ is a non-negative auxiliary function and 
$R(\eta-\zeta)$ is a non-negative normalized kernel function.

Eqs. (12)-(14) define the model used to decompose a long-slit
spectrum of an inhomogeneous background with superposed point-sources 
In fitting
this model to an observed spectrum $\tilde{\phi}(\lambda,y)$, the unknowns
to be estimated are the spectra $f_{n}(\lambda)$ and the
auxiliary function $\chi(\zeta,\lambda)$. The spatially-varying spectrum of
the
background $F(\lambda,\eta)$ is not a basic unknown but is obtained
incidentally from the auxiliary function
$\chi(\zeta,\lambda)$ via Eq.(14). 

As in Sec. 3, the point-spread function $P(y-\eta,\lambda)$ and the
resolution kernel $R$ are user-supplied, thus treated as known functions
in the iterative fitting procedure described below, and the 
centroids $y_n$ of the point source spectra also remain fixed.

\subsection{Iterative technique}

We now estimate the unknown functions $f_{n}$ and $\chi$ by
iteratively improving the fit of
$\phi$ to the observed spectrum $\tilde{\phi}$.
This iterative technique, which is again based on the RL algorithm,
repeatedly sweeps through the spectrum improving both the spectra 
$f_{n}(\lambda)$ and the auxiliary function $\chi(\zeta,\lambda)$ until
convergence is achieved.  

The scheme is initiated with $f_{n}(\lambda) =1$ for all $n$
and with $\chi(\zeta,\lambda)= 1$. In addition, the normalized effective
broadening function
\begin{equation}
 Q(y-\zeta,\lambda) = \int\!\!  P(y-\eta,\lambda) R(\eta-  \zeta) d \eta
\end{equation}
that maps $\chi(\zeta,\lambda)$ into $\omega(\lambda,y)$ - see Eqs.
(13) and (14) - is computed and stored for use during the iterations.

At each iteration, the following calculations are made at every
discrete $\lambda$ in the observed spectrum:

1) The $y$-dependence of the broadened background
$\omega(\lambda,y)$ is computed by
convolving the current estimate of $\chi(\zeta,\lambda)$ with 
$Q(y-\zeta,\lambda)$.

2) The $y$-dependence of $\phi(\lambda,y)$
is computed from Eq.(12) using the current estimates of the amplitudes
$f_{n}$.

3) Improved point-source amplitudes $f_{n}$ are obtained by applying
the correction factors 
\begin{equation}
 c_{n}=\int \frac{\tilde{\phi}(y,\lambda)}{\phi(y,\lambda)}P(y-y_{n},\lambda)dy
\end{equation}

4) The auxiliary function $\chi(\zeta,\lambda)$ is improved by applying the
correction factor
\begin{equation}
 C(\zeta,\lambda) = \int \frac{\tilde{\phi}(y,\lambda)}{\phi(y,\lambda)}Q(y-\zeta,\lambda)dy
\end{equation}

\subsubsection{Test of convergence}

When steps 1)-4) have been carried out at all $\lambda$'s,
convergence is tested. The quantities tested are $\delta_{n}$ as previously
defined by Eq.(10) and $\delta_{\chi}$, which is redefined here as 
\begin{equation}
 \delta_{\chi} = \int \int |\Delta \chi| d \zeta d \lambda /
 \int \int \chi d \zeta 
d\lambda
\end{equation}
If {\em any} of the $\delta_{n}$ or
$\delta_{\chi}$ exceeds an appropriately small quantity
$\epsilon$, control returns to step 1) above in order to carry out a further
iteration. 

\subsubsection{Output}

If the above convergence test is satisfied, the reduction is finished
and it remains only to output the quantities of scientific interest. These
are the spectra of the point sources $f_{n}(\lambda)$ and the two-dimensional 
spectrum of the background $F(\lambda, \eta)$. This latter quantity is
computed from the converged auxiliary function $\chi$ via Eq.(14). 

From $F(\lambda, \eta)$, one-dimensional plots of the background
spectrum at different positions $\eta$ along the slit can be displayed.
Alternatively, one-dimensional plots of the background's spatial profile
at different $\lambda$'s can be extracted.    

As in the homogeneous case, the two-dimensional residual spectrum 
$\phi(\lambda,y)-\tilde{\phi}(\lambda,y)$ is also computed and displayed.
If some residuals are statistically
significant, a flawed or incomplete reduction is indicated - see comment b)
below.

\subsection{Comments}

The following comments refer to differences between this 
iteration scheme and that of Sec. 3.

a) Statistical fluctuations in the extracted background spectrum
$F(\lambda,\eta)$
are reduced by an effective weighted co-addition in the spatial direction
using the broadening function $Q(y-\zeta,\lambda)$ given by Eq. (15).  
  
b) The converged model $\phi(\lambda,y)$ obtained with the above
iterative scheme will not necessarily fit the observed spectrum 
$\tilde{\phi}(\lambda,y)$ to within errors. But significant residuals
are less likely than in
the previous case because the restrictive assumption of a spatially-invariant
background spectrum has been dropped. Nevertheless, as previously,  
significant residuals will remain if the user-supplied
broadening function $P(y-y_{n},\lambda)$ is inaccurate, if the second
channel still contains point-sources, or if the background contains
structure or resolved sources with widths narrower than that of $R$.  

\section{Applications}
The two techniques are distinguished by their treatment of the 
extended source. If the spectrum of the extended source changes with
spatial position across the slit, then the inhomogeneous technique
should be applied; if the spectrum of the
extended source is expected not to vary with position
then the homogeneous code is applicable and will give a more robust 
estimation for the extended source with incidental improvements
in the extracted spectra of the point sources.

The capabilities are illustrated by Figure 1 which shows a simulated 
2D spectrum and the results of extraction of the point source spectra 
using the inhomogeneous technique (Sect. 4). The extended source 
consists of two spectral components with differing spatial extents - 
one with a sloping background and the other with line emission. The 
point sources have respectively a continuum with an absorption line 
and a flat continuum with emission lines. The difference in the 
continua of the two point sources is 1.75mag. at the short 
wavelength end of the spectrum. 
The results for the restored spectrum are shown: the match of the 
restored extracted point source spectra (integrated over the 
spatial extent) and the input templates are excellent. 
The difference between the restored long-slit spectrum (point and 
extended sources) and the input shows only random noise and no 
systematics, validating, in practice, the techniques described in
Section 4.

In reality, the spectrum of the PSF (hence called the Slit Spread 
Function SSF) is often not a perfect match to the data because of
instrumental effects and changing seeing (for ground-based data);  
the exact position of the point source(s) (required to shift the 
SSF to the position of a point source) are also not known. Ideally
a PSF star should be placed on the slit together with the target,
or, in the case of mulit-slit instruments, on another slitlet.
However even having a star on the slit is not a guarantee of 
a good SSF. Experiments with ground-based data on a visual binary 
with separation $\leq$ seeing, using an SSF derived from a star 
at the edge of the slit, resulted in systematic residuals
of the restoration which were spectrally varying. Thus the SSF at the
edge of the slit was not identical to that at the centre on account of
optical aberrations in the spectrometer. For HST spectra taken with
STIS, for example, a spectrum of a point source at approximately the
same position on the slit as the source could be used for the SSF. 
However very often suitable stars have lower signal-to-noise than the 
target or were not similarly centred in the slit. 
In addition two HST SSF's may not be identical because of "breathing"
(thermally induced changes in the telescope focal length), particularly
if the exposure times differ greatly between the object and SSF star. 

In the case of space-based or adaptive optics long-slit spectra, 
the SSF can be constructed from an optical model, such as TinyTim for
HST \citep{kris95}. Tiny Tim delivers a monochromatic PSF or a PSF
integrated over a filter passband. Integrating the PSF with a slit 
of the correct size and pixel sampling allows a monochromatic 
long-slit PSF to be formed. Interpolating several such across the 
wavelength range of a long-slit spectrum allows construction of a 
high fidelity SSF image. In order to obtain the utmost 
from the slit spectra to be decomposed, precuations should be taken 
at the time of observation, or even at the time of instrument design 
in the case of a specialized instrument.

\begin{figure}
\figurenum{1}
\epsscale{0.8}
\plotone{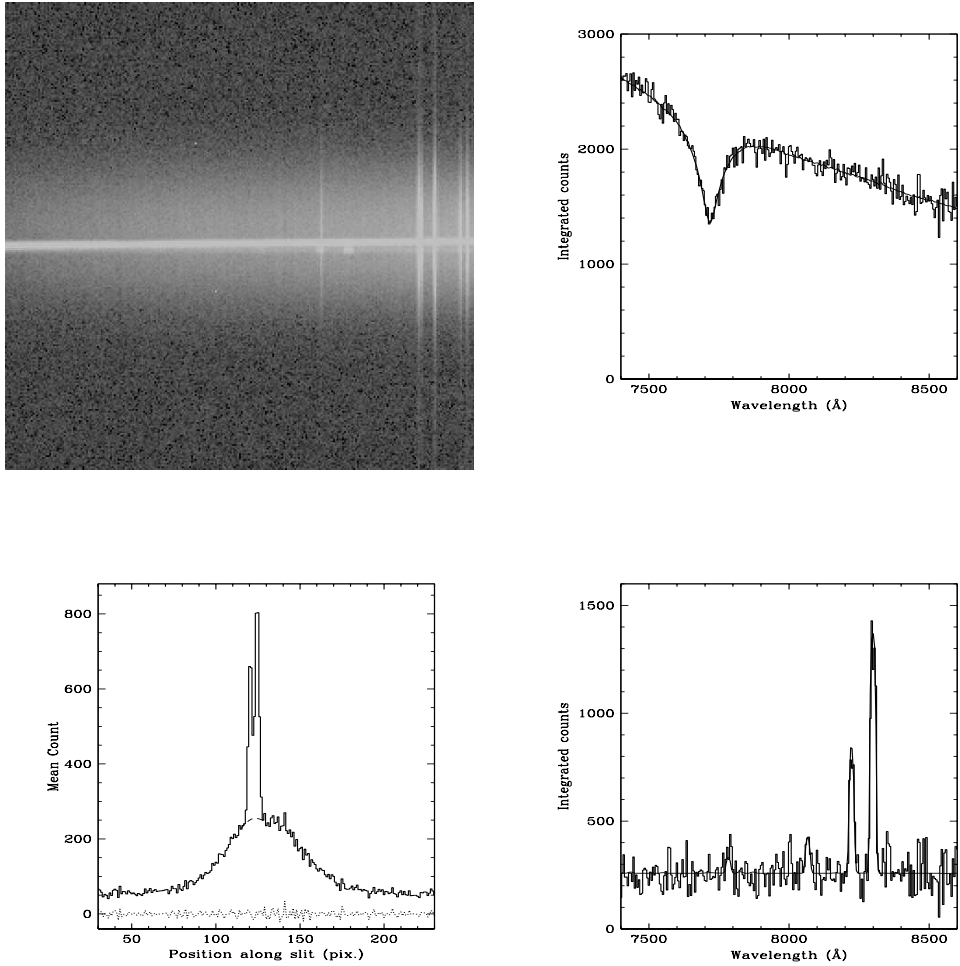}
\caption{The capabilities of the inhomogeneous technique are illustrated
by a simulated 2D spectrum (dispersion in X direction), shown at 
upper left, consisting of two point sources, of FWHM 2.7pixels 
separated by 1.3 times their width. The upper point source has a continuous 
spectrum with a broad absorption line (upper right) and the lower one 
emission lines on a flat continuum (lower right). The extended source is 
a broad Gaussian whose spectrum consists of a sloping continuum
and emission lines with different spatial profiles.
An arbitrary wavelength scale has been applied to the X axis 
and Gaussian distributed random noise was added. The 2D spectrum 
was restored with the inhomogeneous code (Sect. 4) and the restored 
point source spectrum (histogram) is compared with the input spectrum 
(bold line) in the right hand plots for both sources; the spectra
are integrated over the spatial extent of the point sources. A 
converged restoration to 10$^{-4}$ for
the point sources and 10$^{-3}$ for the background was achieved in
91 iterations. At lower left is shown the observed spatial 
profile of the 2D spectrum at the wavelength of the brightest emission line 
($\sim$ 8300\AA), the restored spatial profile of the background 
under the point sources (dashed line) and the residuals on the 
observed-restored profile (dotted line).
}
%
% Figure original in 
% /home/jwalsh/STIS/software/export/demo/rub/
%
\end{figure}

In the following three sub-sections, examples of the application 
of the two techniques to a variety of particular problems are illustrated.

\subsection{Decomposition of AGN spectrum from its galaxy}
Figure 2 shows a long-slit spectrum of the nucleus of the Seyfert I 
galaxy NGC~4151 taken by STIS with the G430M grating (PI:Hutchings; 
Programme ID:7569 ; see \citet{hutc99}). This was actually
taken without a slit. An SSF image was constructed from five TinyTim
PSF's interpolated over the wavelength range of 4830 to 5090\AA\ and with a 
broad slit (0.8$''$). The inhomogeneous code (Sect. 4) 
is obviously necessary for this application since the spectrum of the 
extended source has different emission line contributions across the slit. 
The extracted point source spectrum of the nucleus of NGC~4151 is
shown in the Fig. 2. Here the scientific aim is to extract the spectrum
of the point source nucleus, uncontaminated by the extended line emission.
Several values of the width of the Gaussian smoothing kernel were tried
but the flux of the resulting bright point source showed relatively little 
sensitivity to the kernel width. The other aspect of this extraction would
be to study the emission line spectrum in the close vicinity of the nucleus
from the extended emission with the point source removed. The veracity of
the reconstruction in such a case depends critically on how well the
model SSF matches the SSF at the time of observation.
 
\begin{figure}
\figurenum{2}
\epsscale{0.9}
\plotone{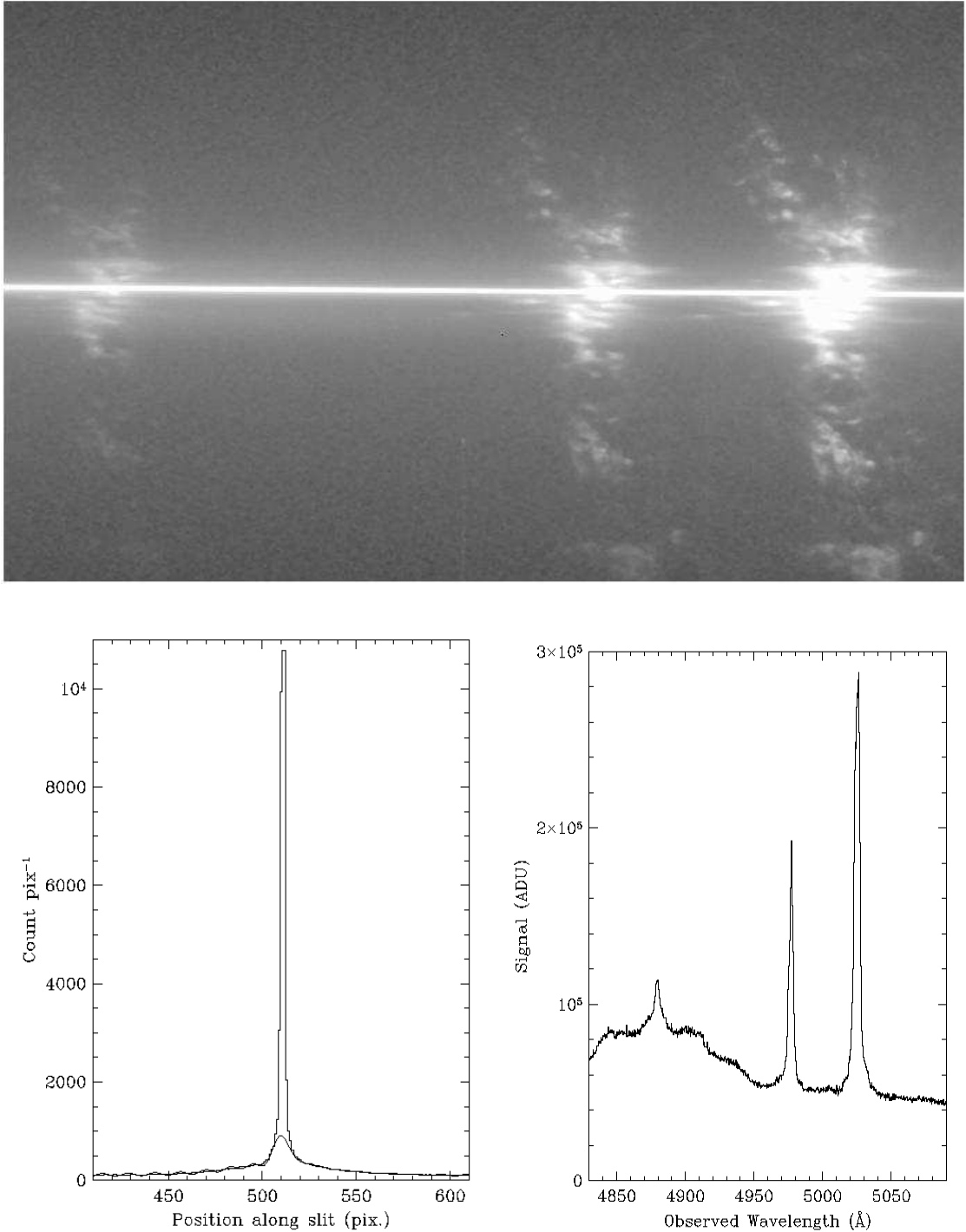}
\caption{A STIS G430M grating longslit CCD spectrum of the Seyfert I 
galaxy NGC~4151 covering the emission lines H$\beta$ and [O~III]
4959 and 5007\AA\ is shown. Restoration with the inhomogeneous code
(background smoothing kernel $\sigma$=4.0 pixels) 
was performed; and 91 iterations
converged to limits of 10$^{-3}$ and 2$\times$10$^{-3}$ for the
point source and background respectively. The PSF spectrum 
was generated from five TinyTim \citep{kris95} PSF's over the wavelength 
range. The spatial profile at a wavelength of 
4930\AA\ (i.e. remote from the emission lines) is
shown at lower left (histogram) and the restored spatial profile as 
a continuous line. Note the extended region of continuum emission
which is distinct from the PSF. The spectrum of the extracted point spread
function (nuclear broad line region) is shown at lower right; the narrow
H$\beta$ and the [O~III] lines have very similar width.
}
%
% Figure original in 
% /home/jwalsh/STIS/N4151/
% test4151_n1.cl
%
\end{figure}

\subsection{Separating extended line emission from a damped QSO 
line trough}
An important class of astronomical problems involve detecting the
spectra of faint extensions around bright point sources, such as the 
underlying galaxy of a QSO, the optical jet emission from a Pre-Main 
Sequence star or an AGN. Such cases mandate the careful subtraction 
of the point source spectrum to reveal the extended spectral features
near the point source. Both techniques presented here are well suited 
to such problems and present a systematic alternative to the ad-hoc
approach of PSF fitting (e.g. \citet{mol00}). Figure 3 shows an 
example of a long-slit spectrum across a broad-absorption line
QSO (Q2059-360). This is a raw spectrum with sky lines included; the 
broad Ly-$\alpha$ absorption line is well seen with a trace of extended 
emission towards its long wavelength edge. The upper, extended object, 
spectrum is of another galaxy. The data was kindly provided by B. 
Leibundgut and was taken with the ESO NTT and EMMI spectrometer.
There was no suitable bright star nearby to
provide the SSF and the spectrum of a spectrophotometric standard
with the same set-up had different seeing. The
spectrum of the BAL QSO itself was used, with the Ly$\alpha$ 
region interpolated and smoothed in the dispersion direction.
The inhomogeneous technique was employed to
separate the Ly$\alpha$ emission source from the BAL continuum
spectrum. The right hand panel of Fig.3 shows the result:
the bright BAL QSO continuum is very effectively subtracted.
The homogeneous technique could also have been
used to model the sky background, which when subtracted from
the spectrum in Fig.3 reveals the extended Ly$\alpha$ emission source.

\begin{figure}
\figurenum{3}
\epsscale{0.9}
\plotone{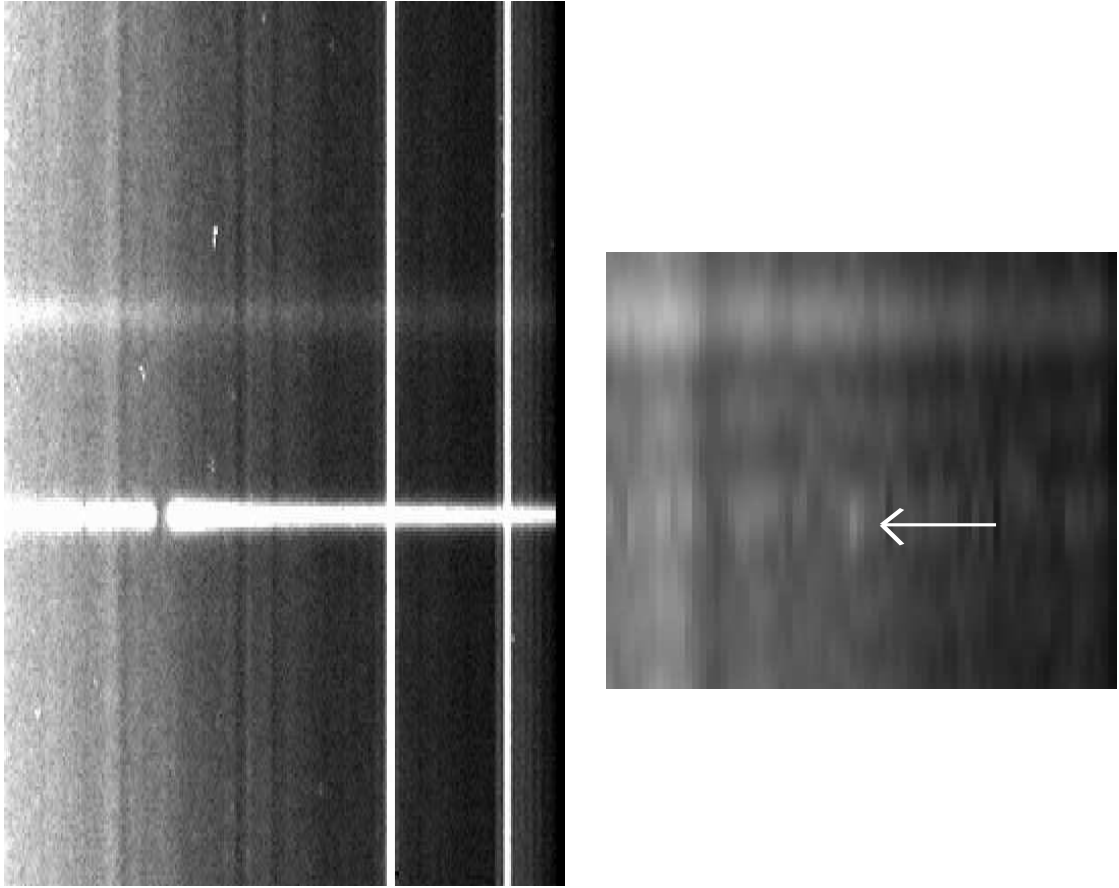}
\caption{A longslit spectrum across the BAL QSO Q2059-360, including 
sky background, is shown in the
left hand image. Weak extended emission to the right (long wavelength) side of the 
Ly$\alpha$ trough can be discerned. The right hand panel shows the
result of application of the inhomogeneous code with the QSO point source 
removed and the background remaining. The spatially extended emission of the 
Ly$\alpha$ emission is arrowed. The smoothing kernel had
a $\sigma$=5.0 pixels and 215 iterations were required to reach convergence
of 0.0003 for point source and background.
}
%
% Figure original in 
% /home/jwalsh/STIS/bruno/
% processa.cl and process3.cl
\end{figure}

\subsection{Extracting many stellar spectra in a globular cluster} 
The last example is a long-slit spectrum of a globular cluster, in this
case NGC~5272 taken with the STIS spectrometer and G430L grating (PI
M. Shara, HST proposal 8226). Figure 4
shows how crowded the spectrum is and that there is essentially no region
which can be used to estimate the background. About 120 point source spectra
could be discerned by eye in this 2D spectrum. Clearly it would be
very difficult, if not impossible, to extract the spectra of all the stars 
on the slit without resort to profile fitting. The positions of all the
stars were determined by fitting Gaussians or simple centroids and a
PSF was provided by the brightest star in the spectrum, suitably cleaned
of its neighbouring spectra. The background is assumed to be composed of
sky, very faint cluster members from orbit mixing and bright sources close to
the slit edge. Assuming this
background does not change its spectral signature across the slit, then
the homogeneous technique should be applied.
The right hand diagram of Fig.4 shows the spatial profile along the slit
and the restored profile of the background with the stars removed.
The first attempts showed point source spectra remaining in the background 
from stars that were closely blended and that were missed in the 
visual identification. Finally the positions of 129 stars were used as 
input to the task and their stellar spectra were recovered. Some high
excursions in the background remain which suggest the presence of stars
omitted from the input catalogue of point sources.

\begin{figure}
\figurenum{4}
\epsscale{0.9}
\plotone{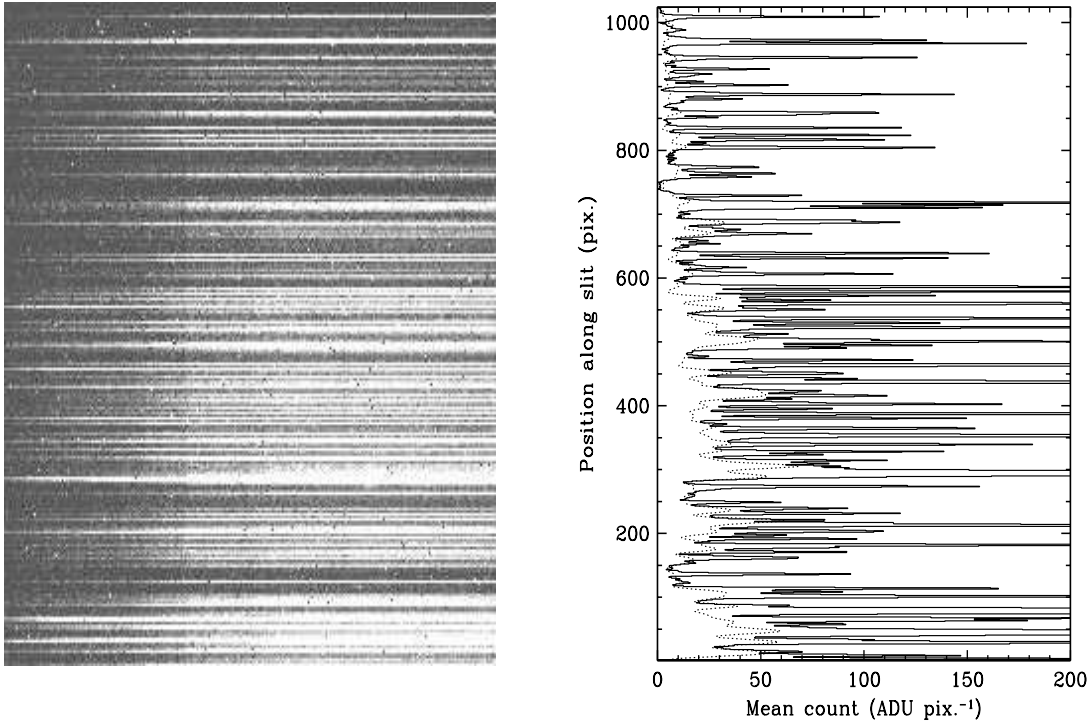}
\caption{The left hand figure shows a STIS G430L spectrum of the 
globular cluster NGC~5272 as an example of a very crowded long-slit
spectrum (wavelength increasing to the right). 129 point source 
spectra were fitted in this spectrum using the homogeneous 
background technique (Sect. 3). The right-hand panel shows the 
spatial profile of the observed 2D spectrum (bold line) and the 
restored background with the stars removed (dashed line) after 50 
iterations to convergence of 10$^{-2}$ for point sources (i.e. 
maximum difference for all point source fluxes between successive 
iterations) and 4$\times$10$^{-3}$ for background. Note the peaks 
in the background (e.g. around channel 300) indicative of further 
point source spectra which were not contained in the input list 
of sources to extract.
}
%
% Figure original in 
% /home/jwalsh/STIS/shara/rub/
%  new1.cl
\end{figure}

\section{Comparison with other techniques}
A variety of other methods for separating the spectra of 
point sources from the spectrum of an extended source have 
recently appeared. \citet{cour00} use an adaption of the
MCS image deconvolution algorithm \citep{mag98}. The required
PSF for the deconvolved image is chosen, and the PSF, which 
should be applied to restore the data to this chosen PSF, is 
applied in the deconvolution. In the long-slit case each 
spectral element is deconvolved independently and three 
functions are minimized for the fit of the point sources, 
the smoothness (on the length scale of the output resolution) 
of the background and the length scale of the correlation in 
the spectral direction (viz. spectral resolution). 
The $\chi^{2}$ minimization requires the assignment of two
Lagrange multipliers. Since the position of the
point-sources is not fitted in the two techniques 
presented here, only one user-specified parameter - the 
width of the smoothing kernel for the extended object spectrum -
is required in comparison. 

  Direct PSF fitting has also been applied to the separation of 
point and extended sources from the spatial component of long-slit 
spectra. \citet{hyne02} presented a method based on the optimal
extraction technique to extract
spatially blended spectra, but prior subtraction of the extended
sources is assumed and not explicitly treated. The maximum entropy method
has been used by \citet{khmi02} in application to overlapping
spectra but here separate $\chi^{2}$ minimization cycles are required
for the spectra of the point sources, the position of the point sources
and the determination of the SSF. Again spatially varying extended 
sources are not covered by the extraction technique. The two R-L
based methods presented here treat the extended source explicitly as
part of the restoration process and thus offer a wider range of
astrophysical applicability. Both these techniques
are based on a well-known image restoration technique. 

For {\em unblended} spectra of point sources, results obtained with the 
techniques described here can be compared with those for the optimal
extraction method of \citet{horn86}. As expected - see Sect. 3.4e, 
the results are then similar, in particular with regard to achieved
signal-to noise.

\section{Software}

The homogeneous and inhomogeneous background extraction codes are 
included in the STECF IRAF\footnote{IRAF is distributed by the 
National Optical Astronomy Observatories, operated
by the Association of Universities for Research in Astronomy, Inc.,    
under contract to the National Science Foundation of the United States.} 
layered package `specres' to extract spectra from longslit or 
2-D spectra with an a priori SSF. The input data are the 2D spectrum 
and the SSF and a table listing the positions of the point sources. 
The output products are the extracted spectra of the point sources and 
the restored background image, either with or without the point sources 
included. A task is also provided in the package 
to produce an SSF image from sets of (2-D spatial) PSF's at different 
wavelengths, simulating
the spatial profile produced by placing a long-slit over a point source.
Further information and help pages for the routines are available at:
http://www.stecf.org/~jwalsh/specres/
The algorithms have been made flexible, can deal with slightly tilted 
spectra, spatially subsampled SSF's, and allow the position of point 
sources to be refined by
cross-correlation. A separate table file is output for each extracted point 
source. Data quality of the input 2D spectrum is considered in order to 
neglect bad pixels and, if statistical errors are available, then multiple 
trials can be performed to determine the error estimates on the restored spectra.

\section{Conclusions}
Two iterative techniques, based on the Richardson-Lucy algorithm, 
have been presented for decomposing a long-slit spectrum into its 
constituent parts, namely into the spectra of the point sources and the 
spectrum of the background. These techniques are applicable to complex 
fields where standard extraction routines fail or perform poorly. A 
variety of tests, with real and simulated data, confirm the ability of 
these techniques to provide effective reductions for complex fields.

\acknowledgments
We would like to thank Richard Hook for allowing us to use his IRAF `cplucy'
code as a basis for the Lucy-Hook algorithms described here and for generous
advice on implementation of restoration methods. We also thank Ian
Howarth and Bruno Leibundgut for providing their ground based spectra
for testing the techniques.

\end{document}